\documentclass{article}
\usepackage{amssymb}
\usepackage{amsthm}
\usepackage{amsmath}
\usepackage{graphicx}

\title{The Fourier-Malliavin Volatility (FMVol)\\MATLAB$^{\circledR}$ library}
\author{Simona Sanfelici (simona.sanfelici@unipr.it),\\  Giacomo Toscano (giacomo.toscano@unifi.it) }
\date{\today}

\begin{document}
 
 \maketitle
 
\begin{abstract}
This paper presents the Fourier-Malliavin Volatility (FMVol) estimation library for MATLAB$^{\circledR}$. This library includes functions that implement Fourier-Malliavin estimators (see \cite{MM02,MM09}) of the volatility and co-volatility of continuous stochastic volatility processes and second-order quantities, like the quarticity (the squared volatility), the volatility of volatility and the leverage (the covariance between changes in the process and changes in its volatility).
The Fourier-Malliavin method is fully non-parametric, does not require equally-spaced observations and is robust to measurement errors, or noise, without any preliminary bias correction or pre-treatment of the observations. Further, in its multivariate version, it is intrinsically robust to irregular and asynchronous sampling. Although originally introduced for a specific application in financial econometrics, namely the estimation of asset volatilities, the Fourier-Malliavin method is a general method that can be applied whenever one is interested in reconstructing the latent volatility and second-order quantities of a continuous stochastic volatility process from discrete observations.
\end{abstract}


\section{Introduction} \label{sect:Introduction}

The accurate estimation of the latent diffusion component of a stochastic process, which gauges the degree of unpredictability, or randomness, of its trajectories, represents a relevant issue in several fields. For instance, in high-frequency financial econometrics, the price of an asset  (e.g., a stock, a bond, a currency)  is typically modeled as an It\^o stochastic differential equation (see, e.g., Chapter 1 in \cite{aj}) and the focus is on the estimation of the diffusion coefficient thereof, which is commonly referred to as the $volatility$ process\footnote{Throughout this paper, we will use the terms $volatility$ and $variance$ as synonyms, as is common practice in the literature on financial econometrics.}. The availability of efficient estimates of asset volatility paths is indeed fundamental for many financial applications, e.g., portfolio allocation,  market risk management and derivatives pricing. However, the issue of estimating the volatility of a random process is not exclusive to financial econometrics but is relevant also for applications in other fields, such as, for instance, climate change studies, computational biology, and medicine (see, e.g., \cite{MUMENTHALER2021102286}, \cite{fan2009bio}, \cite{ annsurg2004} and \cite{papa2013}, respectively).  

The Fourier-Malliavin method, originally introduced in \cite{MM02} and \cite{MM09}, represents an efficient and easily implementable tool to address this task in a multivariate setting.  As detailed in Section \ref{sect:method}, the method involves three steps. Firstly, one computes the Fourier coefficients of the increments of the variables of interest, e.g., asset price returns. Then, these coefficients are exploited to reconstruct those of the latent co-volatility matrix, using a convolution formula. Finally, co-volatility paths are obtained from the co-volatility coefficients via the Fourier-Fejer inversion formula. 

The possibility of recovering the Fourier coefficients of the latent co-volatility matrix from those of the increments of the observed process may not come as a surprise if one recalls the notion of
\textit{Fourier transform}, an integral transform that allows the representation of a periodic function as a linear combination of projections on a trigonometric basis, termed the \textit{frequency-domain representation}.  Specifically, the term Fourier transform refers to both the frequency-domain representation and the mathematical operation that links the latter to a function of time. The Fourier transform represents a flexible tool that has been widely used for applications in different fields, including engineering, physics and also finance. Indeed, operations performed in one domain (time or frequency) have corresponding operations in the other domain that may sometimes be easier to perform. For example, the operation of convolution in the time domain corresponds to multiplication in the frequency domain and vice versa. Therefore, one may apply the Fourier transform to a given function, perform the desired operations more easily in the frequency domain, and finally transform the result back to the time domain by Fourier inversion. For further details, see, e.g., \cite{Bloom}, \cite{Hannan}, \cite{Pries}. 
 
In this paper, we introduce the Fourier-Malliavin Volatility (FMVol) estimation library for MATLAB$^{\circledR}$. The primary function of this library allows one to estimate the co-volatility matrix of a 2-dimensional stochastic process from discrete observations of the latter. Further, the library also yields univariate estimates of the quarticity (i.e the squared volatility), the volatility of the volatility process and the leverage (namely, the covariance between changes in the observable variable of interest and changes in its latent volatility)\footnote{The term leverage comes from the financial economics literature, where the (usually negative) covariance between asset price increments and volatility increments is typically described as a result of the `leverage effect'. See, e.g., \cite{CHRISTIE1982407} for an economic explanation of this empirical stylized fact.}.  The Fourier coefficients of the volatility of volatility and the leverage are obtained through an iteration of the convolution formula  (see Section \ref{sec:iter_conv}). The Fourier coefficients of the quarticity, instead, are obtained through a product formula (see Section \ref{sec:prod}). Note that the library produces both instantaneous (spot) estimates, that is, estimates of trajectories on a discrete grid, and integrated estimates over the observation period.

The Fourier-Malliavin method is intrinsically robust to asynchronous and irregular sampling and to the presence of noise in the observations, due for instance to measurement errors. In financial applications, such measurement errors are typically due to the presence of market microstructure (see Section \ref{sec:Simul}). Moreover, its computational stability is ensured by the fact that it is based only on integration procedures. Indeed, alternative methods for the estimation of the covariance path are typically based on numerical differentiation (see Chapter 8 of \cite{aj}), which is more vulnerable to numerical instabilities. It is worth stressing that, even though the Fourier method has been introduced to address a specific issue in financial econometrics, namely the estimation of asset co-volatilities with high-frequency prices, its applicability is more general, that is, the method can be applied whenever one is interested in estimating the latent diffusion component of a (multi-dimensional) stochastic process. An application to atmospheric temperature data is discussed in Section \ref{Weather}.

The outline of the paper is as follows. Section \ref{sect:FourierTransform} recalls the notion of Fourier transform, while Section \ref{sect:method} illustrates the Fourier-Malliavin estimation method. The FMVol library is introduced in Section \ref{sect:toolbox} and Section \ref{sec:emp} contains two empirical applications of the latter, with financial and weather data, respectively. Finally, Section \ref{sec:concl} concludes.

 \section{The Fourier transform and Fejer's convergence theorem}\label{sect:FourierTransform} 

Given a function $f$ defined and integrable on $[0,T]$, for any integer $k$, the $k$-th Fourier coefficient is defined  as
\begin{equation}
\label{FourierDefini}
{\cal F}(f)(k) := {1\over {T}} \int_0^{T} f(t) e^{-{\rm i}\frac{2\pi}{T}kt}dt,
\end{equation}
where ${\rm i}= \sqrt{-1}$. When the independent variable $t$ represents time, the transform variable $k$ represents the frequency. Accordingly, ${\cal F}(f)(k)$ is also called the Fourier transform of $f$ at frequency $k$. 
 Evaluating ${\cal F}(f)(k)$ for all values of $k$ produces the frequency-domain function.  For example, if time is measured in seconds, then the frequency is measured in hertz.

By the \emph{Fejer theorem}, if the function $f$ is continuous on $[0,T]$, the trigonometric series
\begin{equation}
\label{fejerRecostr}
\sum_{|k|\leq N} \left(1-{|k|\over N} \right) \ {\cal F}(f)(k) \ e^{{\rm i}\frac{2\pi}{T}kt}
\end{equation}
converges uniformly and in mean square to $f(t)$ on $[0,T]$ as $N\to \infty$, see, e.g., \cite{M95}.

Finally, we define 
\begin{equation}
\label{FourierDefini2}
{\cal F}(df)(k):= ~ {\frac{1}{T}}\int _0^{2\pi} e^{-{\rm i}\frac{2\pi}{T}kt }\, df(t)\, .
\end{equation}
Using integration by parts, one obtains that
\begin{equation}
\label{IntByParts}
{\cal F}(df)(k)= {\rm i} k{2\pi \over T} {\cal F}(f)(k) + \frac{f(T)-f(0)}{T}.
\end{equation}

\section{The Fourier-Malliavin estimation method}\label{sect:method}

We describe the Fourier-Malliavin estimation method, originally introduced in \cite{MM02}, \cite{MM09}. The building block of the method is the {\it convolution formula}, which yields an estimator of the Fourier coefficients of the co-variation of two processes. Firstly, we present this formula in the case of a $d$-dimensional diffusion process. Then we address iterated covariances. 
Finally, we examine the {\it product formula}, which may be needed for estimating auxiliary quantities. An example of auxiliary quantity is provided by the quarticity. The latter appears in the asymptotic error variance of volatility estimators (see, e.g., \cite{LiMaMa}) and thus its estimation is needed for obtaining a feasible Central Limit Theorem.

\subsection{Convolution formula}

Consider the $d$-variate stochastic differential system
\begin{equation}
\label{MultiSDE}
dx^j(t)= b^j(t)dt + \sum_{i=1}^l\sigma^j_i (t) dW^i(t), \ \ \ \ \  x^j(0)=x^j_0 \in \mathbb{R}, \ \ j=1, \ldots , d,
\end{equation}
where $W^1, \ldots , W^l$ are $l$ independent Brownian motions, while, for any $i$ and any $j$,
 $\sigma_i^j$ and $b^j$ are random processes, adapted to the Brownian filtration.



The variance-covariance matrix $\Sigma(t)$, whose entries are equal to
$$\Sigma^{i,j}(t)= \sum_{k=1}^l \sigma^i_k(t)\sigma^j_k(t), \ \ \ \ i,j=1, \ldots , d,$$
is also called {\sl volatility matrix}.

In the case of $d$ dependent Brownian motions, if we let the system of stochastic differential equations read as
\begin{equation}
dx^j(t)= b^j(t)dt + \sigma^j (t) dW^j(t), \ \ \ \ \  x^j(0)=x^j_0 \in \mathbb{R}, \ \ \ \ j=1, \ldots , d,
\end{equation}
the volatility matrix $\Sigma(t)$ is then given by 
$$\Sigma^{i,j}(t)= \sum_{k=1}^l \sigma^i_k(t)\sigma^j_k(t) \rho^{i,j}(t), \ \ \ \ i,j=1, \ldots , d,$$
where $\rho^{i,j}(t)$ is the instantaneous correlation at time $t$ between the Brownian motions $W^i$ and $W^j$.

From the mathematical point of view, the entries of $\Sigma^{i,j}(t)$ are related to the quadratic co-variation process $\langle x^i,x^j\rangle_t$\footnote{Let $\Pi$ represents a partition of the interval $[0,t]$ and $\|\Pi\|$ represents the mesh size of $\Pi$. The quadratic co-variation between the stochastic processes $X$ and $Y$ is defined as $$\langle X, Y \rangle_t:= \lim_{{\|\Pi\| \to 0}} \sum_{{i=1}}^n \left( X_{t_i} - X_{t_{i-1}} \right) \left( Y_{t_i} - Y_{t_{i-1}} \right),$$ where the limit, if it exists, is defined using convergence in probability.} by the relation
$$d\langle x^i,x^j\rangle_t=\Sigma^{i,j}(t)dt.$$
This remark is at the basis of the Fourier-Malliavin estimation method of covariances.

For any integer $k$, denote by ${\cal F}(dx^i)(k)$ and ${\cal F}(\Sigma^{i,j})(k)$ the Fourier coefficients of the increments of $x^i$ and of the entries of the volatility matrix, respectively (see definitions (\ref{FourierDefini2}) and (\ref{FourierDefini})).
The fundamental result of \cite{MM02} and \cite{MM09} is that for any integer $k$, it is possible to compute the Fourier coefficients ${\cal F}(\Sigma^{i,j})(k)$ of the latent spot covariances $\Sigma^{i,j}(t)$ by means of the Fourier coefficients ${\cal F}(dx^i)(k)$ and ${\cal F}(dx^j)(k)$.
Precisely, for any $i,j=1, \ldots, d$, it holds that
\begin{equation}  \label{FTCP}
{\frac{1}{T}}\,{\cal F}(\Sigma^{i,j})={\cal F}(dx^i) * {\cal F}(dx^j),
\end{equation}
where the {\sl convolution product}  in (\ref{FTCP}) is defined as 
\begin{equation}  \label{FTCP1}
({\cal F}(dx^i) * {\cal F}(dx^j))(k):=~ \lim _{N\to \infty }{\frac{1}{2N+1}}
\sum_{|s|\leq N} {\cal F}(dx^i) (s) {\cal F}(dx^j)(k-s),
\end{equation}\label{eq:lim_convo}
for any $i,j$ and for all integers $k$.
The convergence of the convolution product in (\ref {FTCP1}) is attained in probability.

Once the Fourier coefficients of the volatility matrix have been computed,
it is possible to reconstruct the spot volatility matrix $\Sigma(t)$ using the Fourier-Fejer inversion formula given in (\ref{fejerRecostr}).
Specifically, if the volatility matrix has continuous paths, namely the functions $t \to \Sigma^{i,j}(t)$ are continuous{\footnote {If $\Sigma^{i,j}(t)$ has cadlag paths, then the limit in (\ref{PSI}) gives $(\Sigma^{i,j}(t)+\Sigma^{i,j}(t^-))/2$, see, e.g., \cite{M95}. }}, then the Fourier-Fejer summation gives, almost surely, that
\begin{equation}
\label{PSI}
\lim_{M\to \infty} \sum_{|k|< M}\left(1-{|k|\over M}\right) {\cal F}(\Sigma^{i,j})(k) \, e^{{\rm i}\frac{2\pi}{T}kt}= \Sigma^{i,j}(t)\, , \ \ \ \ \hbox{ for all } t \in (0,T),
\end{equation}
for $i,j=1,\ldots, d$.

When the diffusion processes are known by discrete observations, we have to define discrete analogues of the quantities introduced above. Without loss of generality, let us consider the case of two processes, observed on the discrete grids $\{ 0=t^j_0< t^j_{1} < \ldots < t^j_{n_j}=T \}$, $j=1,2$. It is worth noting that we allow for irregularly spaced and asynchronous observation times.

For any integer $k$, $|k|\leq 2N$, let us define the discrete Fourier transform of the increment of $x^i$ as
\begin{equation}
\label{FOUMULTcap2} c_k(dx^i_{n_i}):= {1\over {T}} \sum_{l=0}^{n_i-1} e^{-{\rm i}\frac{2\pi}{T}kt^i_l}\delta_{l}(x^j),
\end{equation}
where $\delta_{l}(x^i):=x^i(t^i_{l+1})-x^i(t^i_l)$, \ $l=0, \ldots, n_i-1$, $i=1,2$. For any $|k|\leq N$ and $i,j=1,2$, let us consider the discrete analogue of the convolution in (\ref{FTCP1}), that is,
$$
{1\over {2N+1}} \sum_{|s|\leq N} c_{s}(dx^i_{n_i})c_{k-s}(dx^j_{n_j}).
$$
Given the identity in (\ref{FTCP}), the previous term, when multiplied by $T$, provides an estimate of the $k$-th Fourier coefficient of $\Sigma^{i,j}$. Therefore, we define
\begin{equation}
\label{CONVMULT1}
c_k(\Sigma^{i,j}_{n_i,n_j,N}):={T\over {2N+1}} \sum_{|s|\leq N} c_{s}(dx^i_{n_i})c_{k-s}(dx^j_{n_j}).
\end{equation}
Finally, the random function of time
\begin{equation}
\label{SMIM1}
\widehat\Sigma_{n_i,n_j,N,M}^{i,j} (t):= \sum_{|k|\leq M} \left(1-{|k|\over M}\right)c_k(\Sigma^{i,j}_{n_i,n_j,N}) \, e^{{\rm i}\frac{2\pi}{T}k t}, \,\,\, t \in (0,T),
\end{equation}
will be called the {\sl Fourier-Malliavin estimator} of the instantaneous volatility matrix $\Sigma^{i,j}(t)$.
The choice of suitable values for the cutting frequencies $N$ and $M$ is of the utmost importance to achieve accurate estimates from (potentially noise-affected) sample observations. We refer the reader to the references in the bibliography for that issue.

The Fourier estimator is consistent in the in-fill asymptotics sense, that is, if $\rho^{1,2}:=\max \left(\max_{1,\ldots,n_1}|t^1_i-t^1_{i-1}|,\max_{1,\ldots,n_1}|t^2_i-t^2_{i-1}|\right)$ tends to zero (or, equivalently, the sampling frequency tends to infinity), and if suitable assumptions on the cutting frequencies $N$ and $M$ are satisfied, then the estimator converges in probability to $\Sigma^{i,j}(t)$, $i,j=1,2$. We refer the reader to \cite{MM02}, \cite{MM09}, \cite{plos}, \cite{mancmar} for details on asymptotic results.
The finite sample properties of the Fourier-Malliavin estimator have been studied in \cite{MS2008}, \cite{MS2011} and \cite{mancmar}.

Additionally, based on (\ref{CONVMULT1}), a consistent estimator of the \textit{integrated volatility matrix} $\int_0^{T} \Sigma^{i,j}(t) dt$ is given by
$$\widehat{\Sigma}^{i,j}_{n_i,n_j,N}:= T \ c_0(\Sigma^{i,j}_{n_i,n_j,N})= {T^2\over {2N+1}} \sum_{|s|\leq N} c_{s}(dx^i_{n_i})c_{-s}(dx^j_{n_j}).$$

Finally, some comments are in order.
First, note that we consider inference problems that are defined on a finite time horizon $[0,T]$ and are based on in-fill asymptotics. 
Such inference problems differ from those that one usually encounters in time-series analysis, where the sampling interval is fixed and $T$ goes
to infinity. Operating on a finite horizon allows us to handle samples from data-generating processes that fail to satisfy stationarity or ergodic properties, which are crucial for long-span ($T \to \infty$) asymptotics. Moreover, our inference problems are also characterized by the fact that we can observe only a single path of the processes of interest and thus we estimate path-wise realized co-volatility trajectories.


\subsection{Iterated convolutions}\label{sec:iter_conv}

The knowledge of the Fourier coefficients of the latent instantaneous variances and covariances allows us to handle these processes as observable variables so that we can iterate the convolution formula and compute, for instance, the Fourier coefficients of the co-variation $B^i(t)$ between the stochastic-variance process $\Sigma^{i,i}$ and the process $x^i$, or those of the quadratic variation $C^i(t)$ of the stochastic-variance process $\Sigma^{i,i}$. In the following sections, we will omit the index $i$ to simplify notations.
We refer the reader to the bibliography for a deeper understanding of the different estimators and more specifically to \cite{CuratoSanfelici2015}, \cite{Curato2019}, \cite{CuratoSanfelici2022}, \cite{MancinoToscano}, \cite{tos2022}, \cite{sanfelici} and \cite{ToLiMaMa} for asymptotic results and finite-sample properties.

Indeed, formula (\ref{FTCP}) can be generalized to estimate the Fourier coefficients of the instantaneous co-variation $\varphi$ of any two square-integrable processes $X$ and $Y$ defined on $[0, T]$, that is, we can define
\begin{equation}
\label{GeneralConvolution}
{\cal F}(\varphi)(k):=~ \lim _{N\to \infty }{\frac{T}{2N+1}}
\sum_{|s|\leq N} {\cal F}(dX) (s) {\cal F}(dY)(k-s),
\end{equation}
where the convergence of the convolution product is attained in probability. Then, we can reconstruct the process $\varphi$ as 
\begin{equation}
  \varphi(t)= \lim_{M\to \infty}\sum_{|k|\leq M} \left(1-{|k|\over M}\right){\cal F}(\varphi)(k) \, e^{{\rm i}\frac{2\pi}{T}k t}.  
\end{equation}
However, we stress the fact that the resolution at which we can reconstruct the process reduces at each iteration of the convolution formula (see also Section \ref{subsec:iterated}).

Finally, we remark that this procedure allows us to compute also integrated quantities on $[0,T]$, that is, $\int_0^{T}\varphi(t) dt$ is obtained as
\begin{equation}\label{coeff0}
T \ {\cal F}(\varphi)(0)= \lim_{N\to \infty}{T^2 \over {2N+1}} \sum_{|s|\leq N} {\cal F}(dX) (s) {\cal F}(dY)(-s).
\end{equation}

\subsubsection{Example 1. The leverage}

Consider from now on the diffusion process
$$dx(t)= b(t)dt + \sigma (t) dW(t), \ \ \ \ \  x(0)=x_0 \in \mathbb{R},$$
where the volatility process is driven by the stochastic differential equation
$$d\sigma^2(t)= a(t)dt + \gamma (t) dZ(t), \ \ \ \ \  \sigma^2(0)=\sigma^2_0 > 0.$$
The Brownian motions $W$ and $Z$ may be correlated, with instantaneous correlation $\rho$.
The processes $a(\cdot)$ and $\gamma(\cdot)$ are the drift and volatility of the variance process.

We use the formula in (\ref{GeneralConvolution}) to estimate the \textit{leverage}, i.e. the co-variation $B(t):=\rho \sigma(t)\gamma(t)$ between $\sigma^2$ and the process $x$.
We assume that the diffusion process $x$ is known by $n$ discrete observations. The estimator of the $k$-th Fourier coefficients of $B(t)$ is defined by
\begin{equation}
	\label{coeffB}
	c_k(B_{n,N,M}):= {T \over {2M+1}} \sum_{|j|\leq M} c_j(dx_{n})c_{k-j}(d\sigma^2_{{n},N}),
\end{equation}
where $c_j(d\sigma^2_{{n},N})$  is computed as\footnote{This formula derives from equation (\ref{IntByParts}). Note that the summand $f(T)-f(0)$ can be neglected in the construction of the estimator (\ref{coeffB}), see \cite{Curato2019} for further details.}
\begin{equation}\label{coeffDS}
c_j(d\sigma^2_{{n},N})={\rm i} \, j\, \frac{2\pi}{T} c_j(\sigma^2_{{n},N}).
\end{equation}
Note that the estimator in (\ref{coeffB}) ultimately depends only on the Fourier coefficients $c_j(dx_{n})$, which are computed from sample observations. In fact, the Fourier coefficients of the variance, $c_j(\sigma^2_{n,N})$, have been estimated in the previous step via (\ref{CONVMULT1}) with $j=i$, which depends only on $c_j(dx_{n})$.

The leverage process can therefore be estimated via
\begin{equation}
\widehat B_{n,N,M,L}(t):= \sum_{|k|\leq L} \left(1-{|k|\over L}\right)c_k(B_{n,N,M}) \, e^{{\rm i}\frac{2\pi}{T}k t}.
\label{spotLEV}\end{equation}

Finally, based on (\ref{coeff0}), a consistent estimator of the \textit{integrated leverage} $\int_0^{T}B(t) dt$ is given by
$$\widehat{B}_{{n},N,M}:= T \ c_0(B_{{n},N,M})= {T^2 \over {M+1}} \sum_{|j|\leq M} {\rm i} j {2\pi \over T} \left(1-{|j|\over M}\right) c_j(dx_{n})c_{-j}(\sigma^2_{{n},N}),$$
where the Fejer kernel has been included to reduce the variance of the estimation error.

\subsubsection{Example 2. The volatility of volatility}

We can use the formula in (\ref{GeneralConvolution}) also to estimate \textit{the volatility of volatility}, i.e. the quadratic variation $C(t)$ of the stochastic variance process $\sigma^2(t)$, given by $\gamma^2(t)$.
We assume again that the diffusion processes are known by $n$ discrete observations. The estimator of the $k$-th Fourier coefficients of $C(t)$ is defined by
\begin{equation}
	\label{coeffC}
	c_k(C_{n,N,M}):= {T \over {2M+1}} \sum_{|j|\leq M} c_j(d\sigma^2_{{n},N})c_{k-j}(d\sigma^2_{{n},N}),
\end{equation}
Based on (\ref{coeffDS}), the estimator in (\ref{coeffC}) depends on the Fourier coefficients of the volatility, $c_j(\sigma^2_{{n},N})$, which in turn (see (\ref{CONVMULT1})) depend only on the Fourier coefficients $c_j(dx_{n})$.

The volatility of volatility process can therefore be estimated by
\begin{equation}
\widehat C_{n,N,M,L}(t):= \sum_{|k|\leq L} \left(1-{|k|\over L}\right)c_k(C_{n,N,M}) \, e^{{\rm i}\frac{2\pi}{T}k t}.
\label{spotVoV}\end{equation}

Finally, from (\ref{coeff0}) a consistent estimator of the \textit{integrated volatility of volatility} $\int_0^{T}C(t) dt$ is given by
$$\widehat{C}_{{n},N,M}:= T \ c_0(C_{{n},N,M})= {T^2 \over {2M+1}} \sum_{|j|\leq M} j^2 \left({2\pi \over T}\right)^2 \left(1-{|j|\over M}\right) c_j(\sigma^2_{{n},N})c_{-j}(\sigma^2_{{n},N}),$$
where the Fejer kernel has again been included to reduce the variance of the estimation error.



\subsection{Product formula}\label{sec:prod}

Besides iterated covariances, the Fourier-Malliavin methodology allows us also to estimate the product of two given processes.
Given two square-integrable processes $X$ and $Y$, defined on $[0,T]$, the Fourier coefficients of their product $XY$ can be obtained by the product formula 
\begin{equation}
{\cal F}(XY)(k)=\lim_{M\to \infty}\sum_{|s|\leq M}{\cal F}(X)(s){\cal F}(Y)(k-s),
\label{PRODUCT}\end{equation}
see \cite{M95}. Once again, the knowledge of the Fourier coefficients of the process $XY$ allows us to reconstruct the latter via the Fourier-Fejer inversion formula.

\subsubsection{Example 3. The quarticity}

The result in (\ref{PRODUCT}) allows estimating the path of relevant statistical quantities such as the so-called {\it quarticity} $\sigma^4(t)$, namely the second power of the squared diffusion coefficient of the process $x$.

Additionally, if one is interested in the integral of the product, then they only need to compute the 0-th Fourier coefficient of the product. For instance, the {\it integrated quarticity} $\int_0^T \sigma^4(t) dt$ can be estimated via
$$ T {\cal F}(\sigma^4)(0)=T \lim_{M\to \infty}\sum_{|s|\leq M}{\cal F}(\sigma^2)(s){\cal F}(\sigma^2)(-s).$$

See \cite{LiMaMa} and \cite{MS2012} for further details on asymptotic results and finite-sample properties of the quarticity estimator.
Finally, we stress that to obtain spot and integrated quarticity estimates via the Fourier-Malliavin method one needs to estimate the Fourier coefficients of the volatility but, nonetheless, the knowledge of the instantaneous volatility path is not required. The same can be said for the Fourier-Malliavin estimators of the leverage and the volatility of volatility.


\section{The library}\label{sect:toolbox}

The MATLAB$^{\circledR}$ codes for the Fourier-Malliavin estimators are available by getting the free add-on  
\begin{center}
    {\bf Flexible Statistics and Data Analysis Toolbox (FSDA)} 
\end{center}
developed by the Department of Economics and Management and the Interdepartmental Centre of Robust Statistics (Ro.S.A.) of the University of Parma, and the Joint Research Centre of the European Commission.

All the .m source codes can be found after installing FSDA from the Mathworks file exchange 

\begin{center}
    https://it.mathworks.com/matlabcentral/fileexchange/72999-fsda.
\end{center}

The associated GitHub repo containing the latest versions of the codes can be found at 

\begin{center}
    https://github.com/UniprJRC/FSDA.
\end{center}

The related MATLAB$^{\circledR}$ documentation is also available at

\begin{center}
    http://rosa.unipr.it/FSDA/index.html,
\end{center}
under the category \textbf{Fourier-Malliavin Volatility (FMVol) estimation}.

The library contains functions to estimate the integrated and instantaneous variance/covariance, leverage, volatility of volatility and quarticity from sample observations of a diffusion process, possibly affected by measurement errors. 

In this section, we describe in detail some of these functions and provide also some examples of their implementation with simulated data from a parametric stochastic volatility model. For a full comprehension of the totality of the routines, we refer the reader to the related MATLAB$^{\circledR}$ documentation and the bibliography cited therein. All functions included in the library are listed in Table \ref{tab:funzioni}.

\begin{table}[htbp]
\scriptsize
    \centering
    \begin{tabular}{l|l}
    \hline
       Function  &  Description \\
       \hline
       \verb|Heston1D.m| & generates a sample trajectory from the  Heston model\\
      \verb|Heston2D.m| & generates a sample trajectory from the bivariate Heston model\\
        \verb|FM_spot_vol.m| & estimates the spot variance of a diffusion process \\
        \verb|FM_spot_cov.m| & estimates the spot covariance of a bi-variate diffusion process \\
      \verb|FM_spot_volvol.m| & estimates the spot volatility of volatility of a stochastic volatility process \\      
         \verb|FM_spot_lev.m| & estimates the spot leverage of a stochastic volatility process \\
         \verb|FM_spot_quart.m| & estimates the spot quarticity of a diffusion process\\
        \verb|FM_int_vol.m| & estimates the integrated variance of a diffusion process \\
         \verb|FM_int_cov.m| & estimates the integrated covariance of a bi-variate diffusion process \\
          \verb|FM_int_volvol.m| & estimates the integrated volatility of volatility of a stochastic volatility process\\
       \verb|FM_int_lev.m| & estimates the integrated leverage of a stochastic volatility process\\
       \verb|FM_int_quart.m| & estimates the integrated quarticity of a diffusion process \\
         \hline \hline
         Supplementary function   &  Description \\
       \hline
        \verb|FE_spot_vol.m| & estimates the instantaneous variance of a diffusion process by the Fourier estimator \\
       \verb|FE_spot_vol_FFT.m| & estimates the spot variance of a diffusion process via the Fourier estimator with the FFT algorithm \\
         \verb|FE_int_vol.m|  &  estimates the spot variance from a diffusion process via the Fourier estimator with Dirichlet kernel  \\
       \verb|FE_int_vol_Fejer.m| & estimates the spot variance from a diffusion process via the Fourier estimator with Fejer kernel \\
       \verb|OptimalCuttingFrequency.m| & computes the optimal cutting frequencies for the Fourier estimator of the integrated variance \\ & in the presence of noise (see \cite{MS2008}).\\
    \end{tabular}
    \caption{List of the primary functions (top) and supplementary functions (bottom) related to the Fourier-Malliavin estimation method that are available in the FMVol library of the \textsc{Flexible Statistics and Data Analysis Toolbox}.}\label{tab:funzioni}
    \label{tab:my_label}
\end{table}

\subsection{Simulation of input data}
We assume our time series data are discrete observations from a bi-variate 
diffusion process. To describe our MATLAB$^{\circledR}$ library, we generate a single trajectory of a bi-variate
version of the well-known stochastic volatility model by \cite{Heston93}:
\begin{eqnarray}
dx^1(t)&=&\left(\mu_1-\frac{1}{2}v^1(t) \right)dt+\sqrt{v^1(t)} dW^1_t \label{Heston1}\\
dx^2(t)&=&\left(\mu_2-\frac{1}{2}v^2(t) \right)dt+\sqrt{v^2(t)} dW^2_t \label{Heston2}\\
dv^1(t)&=&\theta_1(\alpha_1 -v^1(t))dt+\gamma_1\sqrt{v^1(t)} dW^3_t \label{Heston3}\\
dv^2(t)&=&\theta_2(\alpha_2 -v^2(t))dt+\gamma_2\sqrt{v^2(t)} dW^4_t
\label{Heston4}\end{eqnarray}
where $W^1, W^2, W^3, W^4$ are correlated Brownian motions such that $\langle dW_t^k,dW_t^r\rangle= \rho_{k,r} \, dt$.
{The simulation approach allows us to appreciate the quality of our estimates.}
The Appendix describes how to simulate this model by using the MATLAB$^{\circledR}$ functions \verb|Heston1D.m| and \verb|Heston2D.m|, which are available in our library.

For simplicity,  we assume that observations from the bivariate model are synchronous. Accordingly, for any positive integer $n$, we let ${\cal S}_{n}:=\{ 0=t_{0}\leq \cdots \leq t_{n}=T  \}$ denote the set of observation times. 

We simulate also the case when observations are affected by noise (see also \cite{MS2008}), that is, the case when one can only observe
\begin{equation}
    \tilde x^j(t_i)=x^j(t_i)+\eta^j(t_i), \ \ \ i=1,\ldots n, \ j=1,2,
\label{noise}\end{equation}
where the noise terms $\eta^1$ and $\eta^2$ are assumed to be i.i.d. sequences of random variables with mean equal to zero and finite variance. Further, they are assumed to be independent of each other. 


\subsection{Variance and covariance}\label{sec:Simul}


The function
\begin{verbatim}
V_spot=FM_spot_vol(x,t,T,'N',N,'M',M,'tau',tau)
\end{verbatim}
computes the spot volatility of a diffusion process via the Fourier-Malliavin estimator in (\ref{SMIM1}) by using as input a vector of observations \verb|x| and the related observation times \verb|t|, the time horizon \verb|T|, the cutting frequencies \verb|N| and \verb|M| and the vector \verb|tau| of estimation times. If \verb|tau| is not specified, then it is set equal to \verb|0:T/(2*M):T| and is given as an output variable \verb|tau_out|.
When not specified, default values for the cutting frequencies are
$$\verb|N|=\lfloor n/2 \rfloor, \ \verb|M|=\lfloor (n/2)^{0.5} \rfloor.$$
The default choices for these and other cutting frequencies in the FMVol library reflect the rate-efficient conditions in the absence of noise, derived in the previously cited papers. 

Similarly, the function
\begin{verbatim}
C_spot=FM_spot_cov(x1,x2,t1,t2,T,'N',N,'M',M,'tau',tau)
\end{verbatim}
computes the spot covariance of a bivariate diffusion process via the Fourier-Malliavin estimator in (\ref{SMIM1}).
 When not specified, the cutting frequencies are set to
 $$\verb|N|=\lfloor \min(n_1,n_2)/2 \rfloor, \ \verb|M|=\lfloor (\min(n_1,n_2)/2)^{0.5}\rfloor.$$
 If the estimation time vector \verb|tau| is not specified, then it is set equal to \verb|0:T/(2*M):T| and given as an output variable \verb|tau_out|.
 
  

\begin{figure}[h!]
   \centering
    \includegraphics{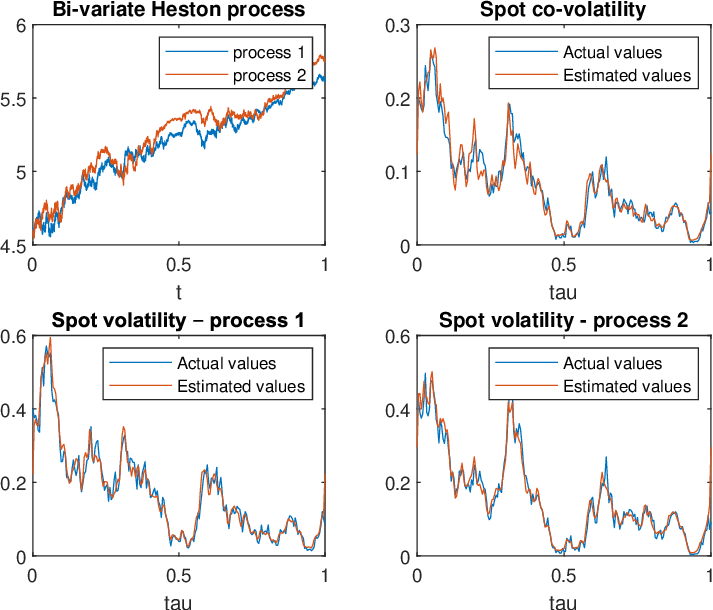}
    \caption{Instantaneous variance and covariance estimation for a bi-variate Heston process.}
    \label{FigVolatility}
\end{figure}

Figure \ref{FigVolatility} shows a comparison between discrete simulated trajectories of the instantaneous variance and covariance processes   
of the model in (\ref{Heston1})-(\ref{Heston4}) and estimates thereof, obtained via the Fourier-Malliavin estimator. The latter has been implemented by using as input the vectors of simulated processes and by selecting default values for \verb|N|, \verb|M| and \verb|tau|. For the simulation, we set \verb|T=1|, \verb|n=23400| and select the following model parameters (see the Appendix):

\begin{center}
     \verb|parameters=[0,0;0.4,0.4;2,2;1,1]|, \verb|Rho=[0.5,-0.5,0,0,-0.5,0.5]|, 
    
    \verb|x0=[log(100)|; \verb|log(100)]|, \verb|V0=[0.4; 0.4]|. 
\end{center}

The MATLAB$^{\circledR}$ random number generator has been initialized by \verb|rng(123456)|. The comparison with the true variance/covariance trajectories shows that the estimation process is very efficient.

\begin{figure}[h!]
   \centering
    \includegraphics{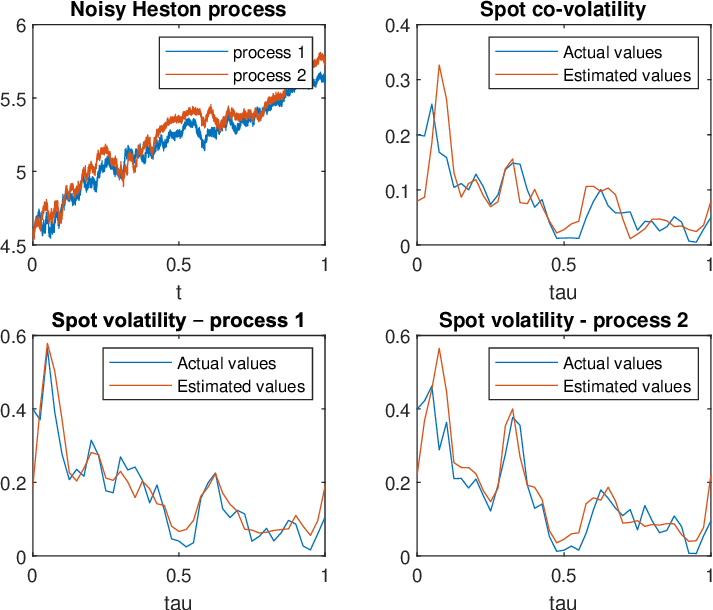}
    \caption{Instantaneous variance and covariance estimation for a noisy bi-variate Heston process.}
    \label{FigVolatility_noise}
\end{figure}

Figure \ref{FigVolatility_noise} is analogous to Figure \ref{FigVolatility}, except for the fact that this time we use vectors of noise-affected observations as inputs. Noise has been simulated according to equation (\ref{noise}), by adding the i.i.d variables $\eta^j(t_i) \sim N(0,\xi_j^2)$, where $\xi_j$ is three times the standard deviation of the increments of the process $x^j$. Based on the finite-sample optimal results in the presence of noise by \cite{MS2008} and \cite{MS2011}, the cutting frequencies have been set equal to \verb|N|$=400$ and \verb|M|$=20$. The resulting equally spaced estimation grid \verb|tau| includes $41$ nodes. The comparison with the true variance/covariance trajectories shows that the estimation process is still very efficient. 


Table \ref{tab:integrate} displays the estimated integrated variances and covariance, compared with their actual values. The estimated values are obtained by calling
\begin{verbatim}
V_int1=FM_int_vol(x1,t1,T)
V_int2=FM_int_vol(x2,t2,T)
C_int=FM_int_cov(x1,x2,t1,t2,T)
\end{verbatim}
with default values of the cutting frequency \verb|N|. The last row of the table displays the estimated values in the noisy case, obtained by setting \verb|N|$=400$. We note that estimates are still very good in the presence of noise, except in the case of process $2$, where the estimate is slightly less efficient for \verb|N|$=400$.

 \begin{table}[ht!]
  \centering
  \begin{tabular}{ llll }
    & {Int. Var. process 1} &  {Int. Var. process 2} &  {Int. Covar.}  \\ 
     \hline
     Actual values & $0.1700$ & $0.1568$ & $0.0805$ \\
     Estimates & $0.1723$ & $0.1594$ & $0.0821$ \\
     Estimates (Noisy data) & $0.1772$ & $0.1764$ & $0.0818$
  \end{tabular}
    \caption{Estimates of integrated variances and covariance and actual values.}\label{tab:integrate}
\end{table}

 \subsection{Leverage, Volatility of Volatility and Quarticity}\label{subsec:iterated}

The estimation of the instantaneous leverage process $B(t)$ by the Fourier estimator in (\ref{spotLEV}) can be performed by calling the function
\begin{verbatim}
L_spot=FM_spot_lev(x,t,T,'N',N,'M',M,'L',L,'tau',tau)
\end{verbatim}
The user must provide as inputs a vector of observations \verb|x| and the related observation times \verb|t|, the time horizon $T$, the cutting frequencies \verb|N|, \verb|M| and \verb|L|, and the vector \verb|tau| of estimation times. If \verb|tau| is not specified, it is set equal to \verb|0:T/(2*L):T| and given as an output variable \verb|tau_out|.
When not specified, default values for the cutting frequencies are
$$\verb|N|=\lfloor n/2\rfloor , \ \verb|M|=\lfloor (n/2)^{0.5}\rfloor, \verb|L|=\lfloor (n/2)^{0.25}\rfloor.$$
Similarly, the estimation of the integrated leverage is performed by calling
\begin{verbatim}
L_int= FM_int_lev(x,t,T,'N',N,'M',M).
\end{verbatim}

\bigskip
Further, the estimation of the instantaneous volatility of volatility process $C(t)$ by the Fourier estimator in (\ref{spotVoV}) can be performed by calling the function
\begin{verbatim}
VV_spot=FM_spot_volvol(x,t,T,'N',N,'M',M,'L',L,'tau',tau). 
\end{verbatim}
As for the inputs, the same considerations apply as for the case of the leverage, with the exception that in the case of the volatility of volatility the default values of the cutting frequencies are\footnote{As shown in \cite{ToLiMaMa}, the default choice of $M=O((n/2)^{0.4})$ guarantees asymptotic unbiasedness for the estimation of the Fourier coefficients of the volatility of volatility in the absence of noise but is not rate-optimal; to obtain a rate-optimal estimator one has to subtract a multiple of the estimated quarticity.}
$$\verb|N|=\lfloor n/2\rfloor , \ \verb|M|=\lfloor (n/2)^{0.4}\rfloor, \verb|L|=\lfloor (n/2)^{0.2}\rfloor.$$
Similarly, the estimation of the integrated volatility of volatility is performed by calling
\begin{verbatim}
VV_int=FM_int_volvol(x,t,T,'N',N 'M', M).
\end{verbatim}

\bigskip
Finally, the estimation of the instantaneous quarticity process $\sigma^4(t)$ by the Fourier estimator can be performed by calling the function
\begin{verbatim}
Q_spot=FM_spot_quart(x,t,T,'N',N,'M',M,'L',L,'tau',tau)
\end{verbatim}
As for the inputs, the same considerations apply as for the case of the leverage. In particular, the default values of the cutting frequencies are the same as for the leverage. Similarly, the estimation of the integrated quarticity is obtained by calling
\begin{verbatim}
Q_int=FM_int_quart(x,t,T,'N',N,'M',M).
\end{verbatim}

\section{Empirical applications}\label{sec:emp}

In this Section, we provide two examples of applications of the FMVol library with empirical data. The first example, which employs high-frequency prices of equity stocks, represents the typical application of the Fourier-Malliavin method. The second example, which involves intraday atmospheric temperature measurements, demonstrates the flexibility of the Fourier-Malliavin method for applications outside the field of financial econometrics.

\subsection{Application to financial data}

\begin{figure}[h!]
   \centering
    \includegraphics[scale=0.2]{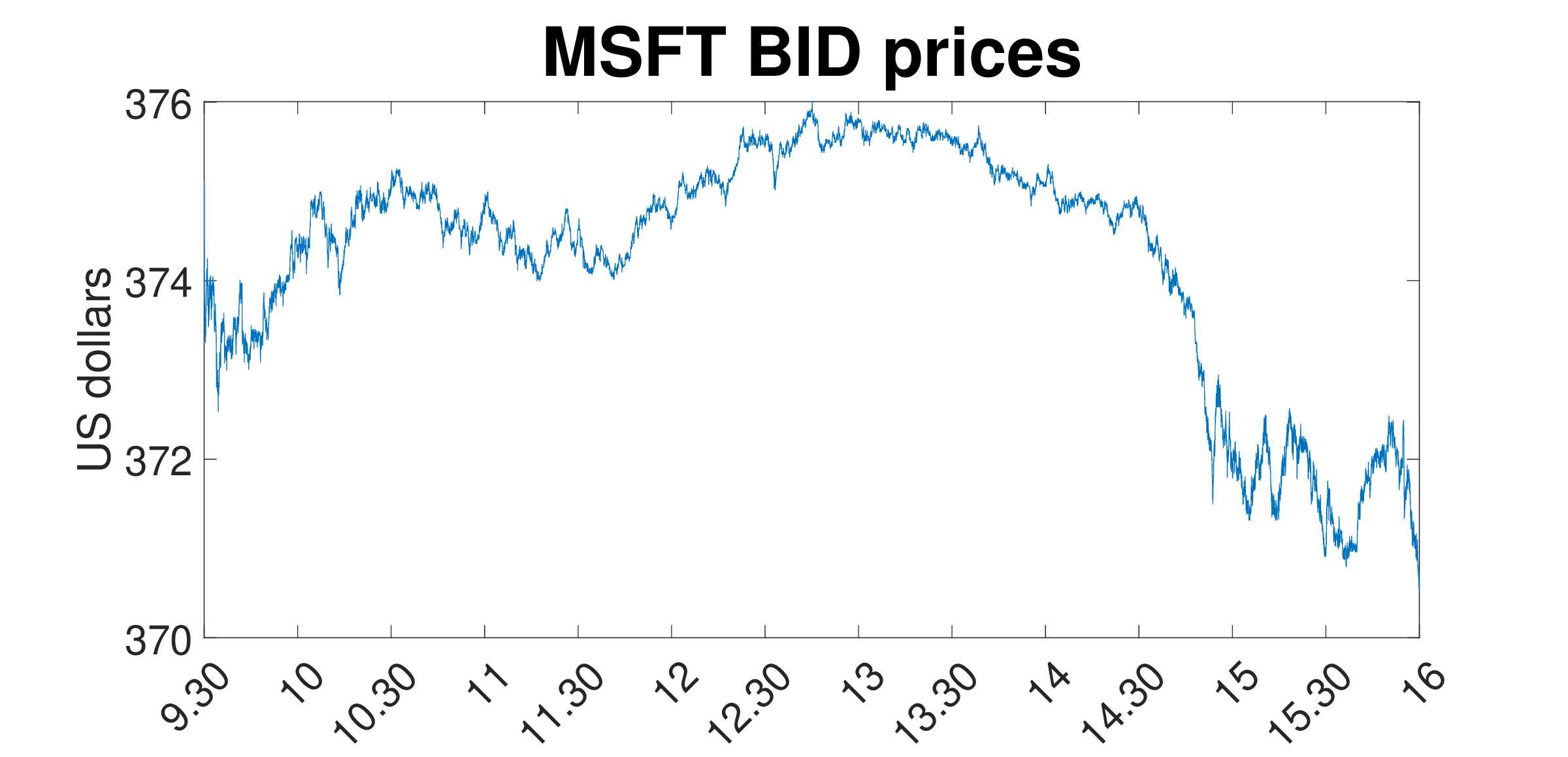}
    \caption{Intraday BID prices of MSFT on December 20, 2023, from 9.30 (market opening) to 16.00 (market closure).}
    \label{fig:MSFT_prices}
\end{figure}

The first application of the FMVol library that we propose employs the series of tick-by-tick BID prices of Microsoft (MSFT) sampled on December 20, 2023. This series has been downloaded from the database of Dukascopy Bank SA\footnote{www.dukascopy.com/swiss/english/marketwatch/historical}. Figure \ref{fig:MSFT_prices} shows the plot of the series. 

The sample size of the series is $n=50151$ and observations are irregularly spaced. Further, observations are affected by measurement error, due to the presence of microstructure noise. 
In this setting,  the consistent estimation of the Fourier coefficients of the volatility requires reducing the cutting frequency $N$, that is, excluding the highest frequencies from the convolution formula in (\ref{CONVMULT1}). The rate-efficient condition for the selection of $N$ in the presence of microstructure noise is found in \cite{mancmar} and is $N=O(n^{0.5})$. Accordingly, using as guidance the simulation study in \cite{mancmar} for the application of the FMVol library, we make the selection: 
 $$ \verb|N|= \lfloor 5n^{0.5} \rfloor, \qquad \verb|M|=\lfloor 0.3 \verb|N|^{0.5}\rfloor, \qquad \verb|L|=\lfloor \verb|M|^{0.5} \rfloor.$$
Further, note that we measure time in days, that is, we set  \verb|T|$=1$, and estimate spot quantities every half hour, that is, we select $\verb|tau|=0:\verb|T|/13:\verb|T|.$

\begin{figure}[h!]
    \centering
    \includegraphics[width=\textwidth]{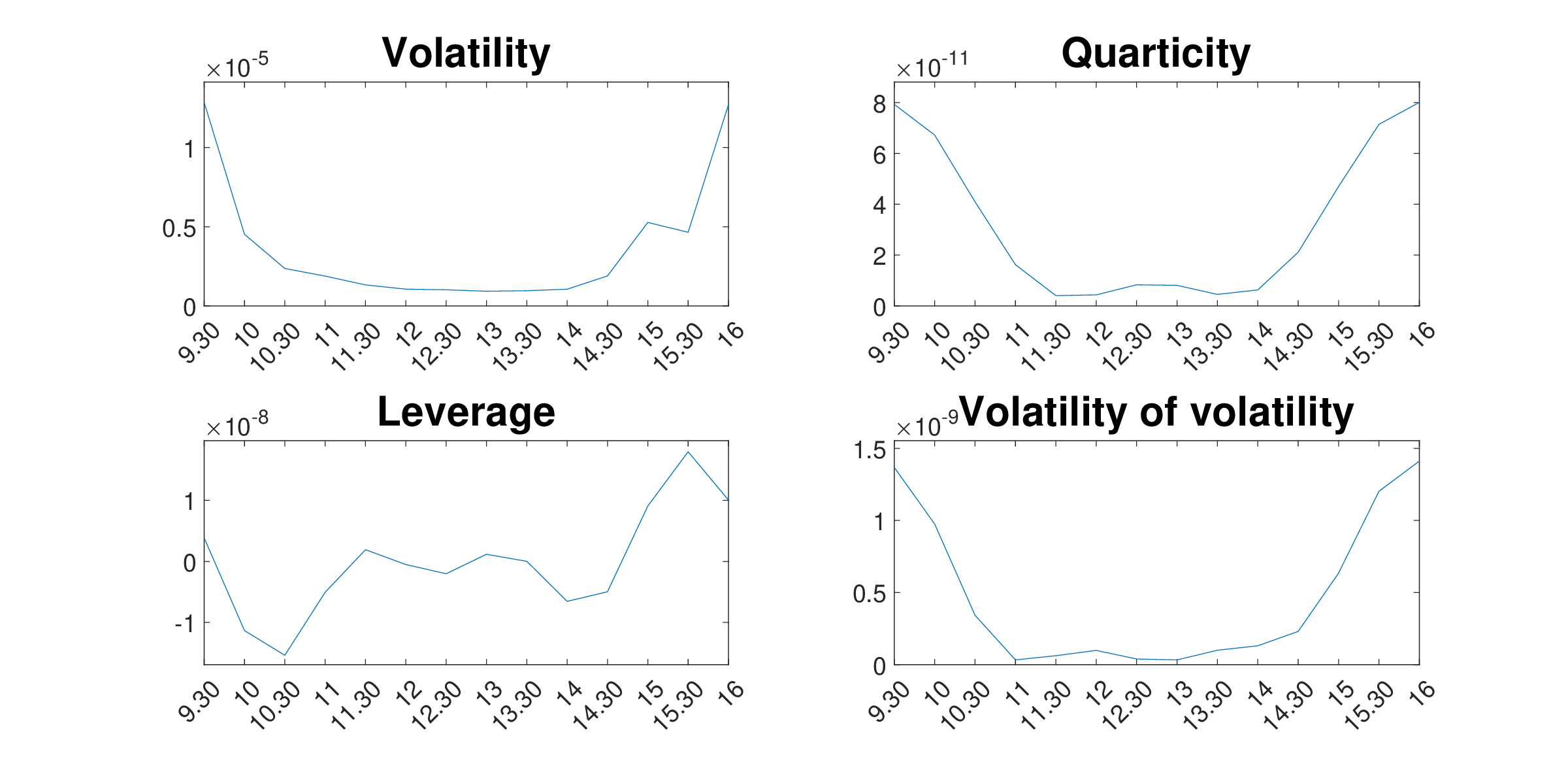}
    \caption{Estimated spot volatility, quarticity, leverage and volatility of volatility for MSFT on December 20, 2023.}
    \label{fig:MSFT_est_paths}
\end{figure}

The results of the application of the FMVol functions for the estimation of spot quantities are displayed in Figure \ref{fig:MSFT_est_paths}, while Table \ref{tab:MSFT_int} resumes the results of the application of the functions for estimating integrated quantities. Figure \ref{fig:MSFT_est_paths} shows that MSFT spot volatility estimates follow the typical U-shape pattern observed in intraday asset volatility series, due to the more intense trading activity near market open and close time. A similar pattern is mechanically reflected in the intraday behavior of spot quarticity estimates. Further, we note that the estimated spot volatility of volatility of MSFT also follows a U-shaped intraday pattern. Finally, we observe that the spot leverage of MSFT is prevalently negative, consistently with the negative correlation between asset returns and their volatilities typically observed on financial markets.  

 \begin{table}[ht!]
  \centering
  \begin{tabular}{ llll }
     {Volatility} &  {Quarticity} &  {Vol. of vol.} & {Leverage} \\ 
     \hline
     $1.087 \cdot 10^{-4}$ & $3.591 \cdot 10^{-8}$ & $2.056 \cdot 10^{-7}$ & $-1.768 \cdot 10^{-6}$ \\
  \end{tabular}
    \caption{Estimates of integrated quantities for MSFT on December 20, 2023.}\label{tab:MSFT_int}
\end{table}

We also apply the FMVol library to estimate the covariance of MSFT BID prices with those of Apple (AAPL) on December 20, 2023\footnote{AAPL BID prices have been also been downloaded from the Dukascopy database.}. The sample size of AAPL BID prices on December 20, 2023 results in $62656$ observations. These observations are also irregularly spaced and, further, non-synchronous with those of MSFT. Accordingly, for the estimation of the covariance between the two stocks we make the following choices of the cutting frequencies.

\begin{figure}[h!]
    \centering
    \includegraphics[scale=0.2]{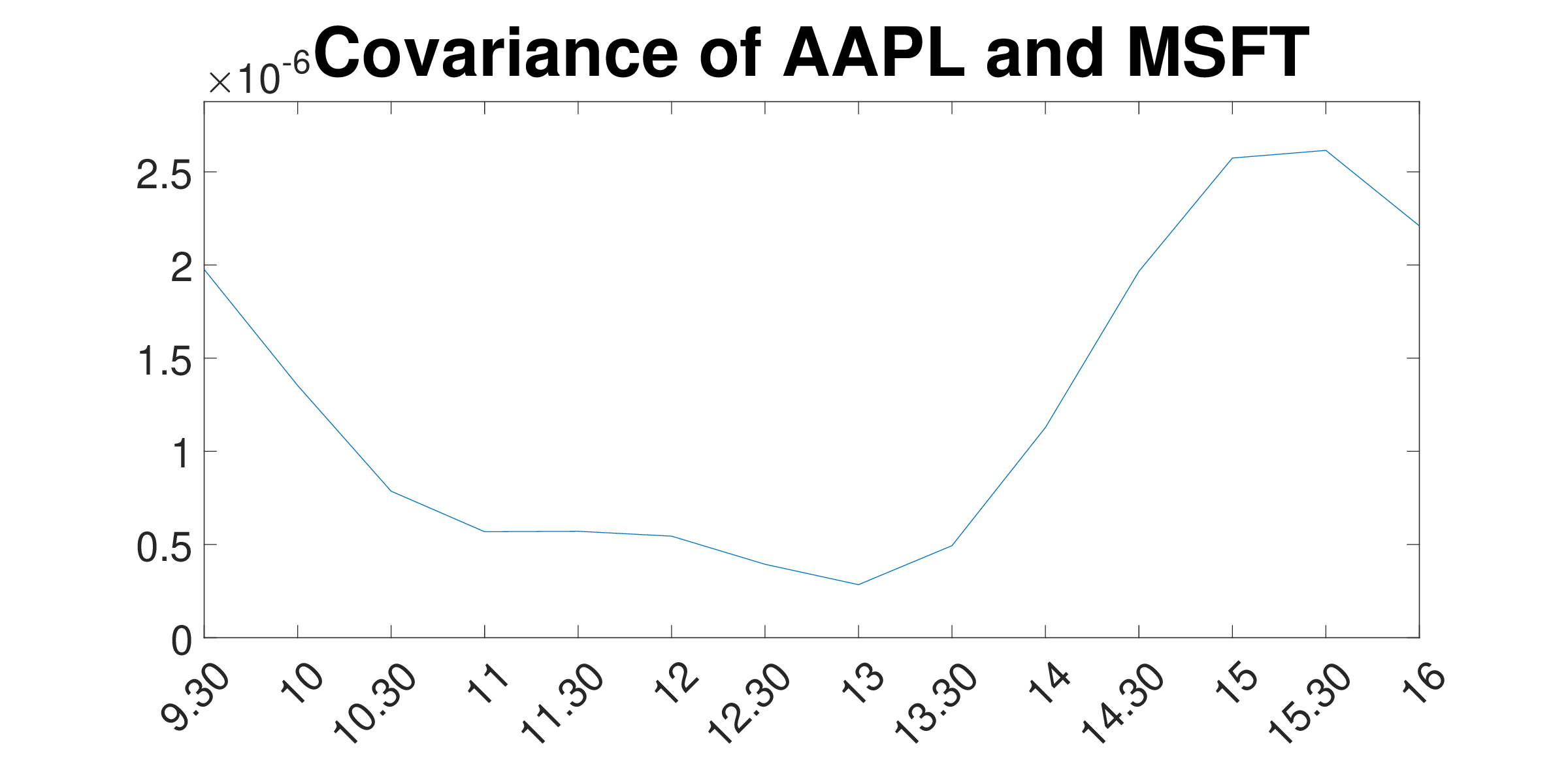 }
    \caption{Estimated spot covariance between MSFT and AAPL on December 20, 2023.}
    \label{fig:cov_est}
\end{figure}

Letting $n_1$ and $n_2$ denote, respectively, the sample size of MSFT and AAPL,  we set $\rho(n)=max\left( max_{0 \le h \le n_1-1}  \left( t^1_{h+1} -t^1_{h}\right),  max_{0 \le h \le n_2-1}  \left( t^2_{h+1} -t^2_{h}\right)\right)$ and select
$$\verb|N| = \lfloor 20 \rho(n)^{-0.5} \rfloor  \qquad \text{and} \qquad \verb|M|=\lfloor  0.3 \verb|N|^{-0.5} \rfloor.$$
As for \verb|T| and \verb|tau|, we maintain the same choice as in the case of the univariate application of the FMVol library with MSFT BID prices. The estimated spot covariance path is shown in Figure \ref{fig:cov_est}. The value of the estimated daily integrated covariance is $3.603 \cdot 10^{-5}$. We note that the spot covariance between the two assets is always positive on December 20, 2023. The positive sign of the spot and integrated covariance can be explained by the fact that both stocks belong to the same industrial sector, namely technology. Further, we note that the spot covariance is higher near market open and close time. 

\subsection{Application to weather data}\label{Weather}

\begin{figure}[h!]
        \centering
        \includegraphics[scale=0.2]{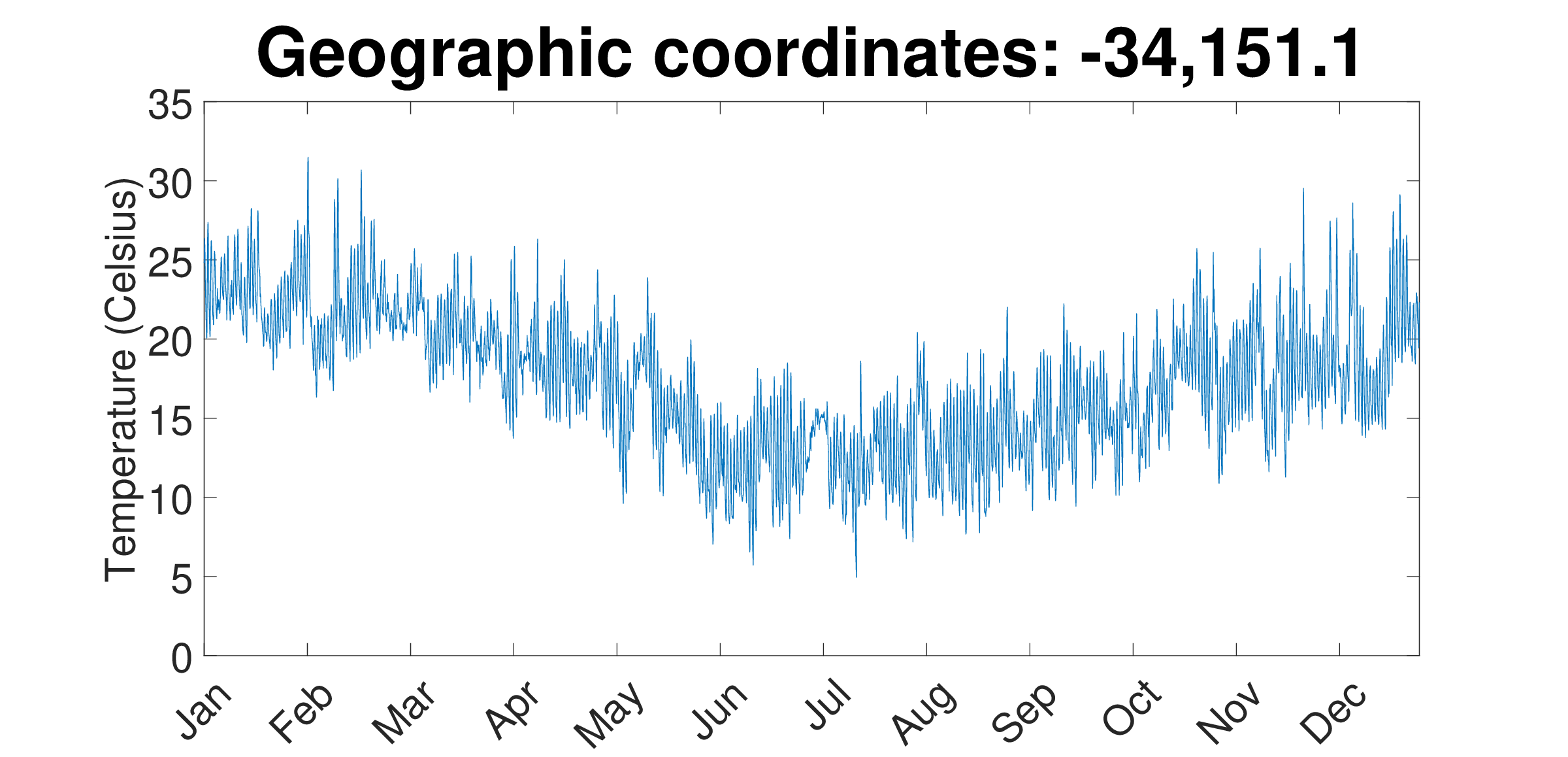}
        \caption{Hourly atmospheric temperature measurements corresponding to the geographic coordinates $-34.0, 151.1$ for the period January 1, 2022 - December 31, 2022.}
        \label{fig:temp_series}
\end{figure}

The second application of the FMVol library that we illustrate considers a time series of atmospheric temperature measurements. 
Specifically, it involves sample intraday atmospheric temperature measurements recorded by a weather station near Sydney, Australia, with geographic coordinates given by $-34.0, 151.1$. This time series, which we downloaded from the Copernicus database\footnote{www.copernicus.eu.}, is sampled at the hourly frequency and covers the period from January 1, 2022, to December 31, 2022, for a total of $8760$ hourly observations. Figure \ref{fig:temp_series} plots the measurement series, expressed in Celsius degrees.
 
As detailed in \cite{mudelsee}, climate time series can be assumed to be a stochastic process composed of a trend (drift) and an i.i.d. noise (diffusion), plus possible outliers (jumps). We note that the effect of the drift on the convolution in (\ref{eq:lim_convo}) is negligible (see lemma 2.2 by \cite{MM09}) and thus, at high frequencies, the presence of a drift component does not impact the finite-sample efficiency of the Fourier-Malliavin method in reconstructing the coefficients of the volatility. However, the time series employed in this example are sampled at the (not so-high) hourly frequency. Hence, we prefer to take a conservative approach and de-trend the series, that is, isolate the diffusion component, in order to avoid a possible finite-sample bias in the estimates of the Fourier coefficients of the volatility. Note that weather time series can incorporate multiple seasonal trends, with different periods. In fact, in our specific instance, the presence of at least two trends can be recognized: a short-span trend related to the repetition of the intraday pattern of temperatures and a long-span trend related to the alternation of summer, autumn, winter and spring throughout the year. 

Specifically, we apply the following procedure to remove trends. Firstly, to eliminate the presence of seasonal trends with multi-week periods, we split the $1$-year atmospheric temperature sample into $73$ sub-samples of the same size, equal to $5$ days. Note that this splitting maintains a sufficiently large number of observations in each sub-sample for the asymptotic theory to be effective, namely $120$. Further, we use the MATLAB$^{\circledR}$ function $trenddecomp$ to find and remove short-span intraday trends from sub-samples. As a result, we obtain a time series of detrended residuals, which represent the diffusion component of the atmospheric temperature series. We use such de-trended residuals as input to the FMVol functions.

For the application of the FMVol functions, we set \verb|T|$=5$, $n=120$ and \verb|tau|$= 0 : 1/6 :$ \verb|T|$-1/6$, and use the default values of the cutting frequencies. 

\begin{figure}[h!]
        \centering
        \includegraphics[width=\textwidth]{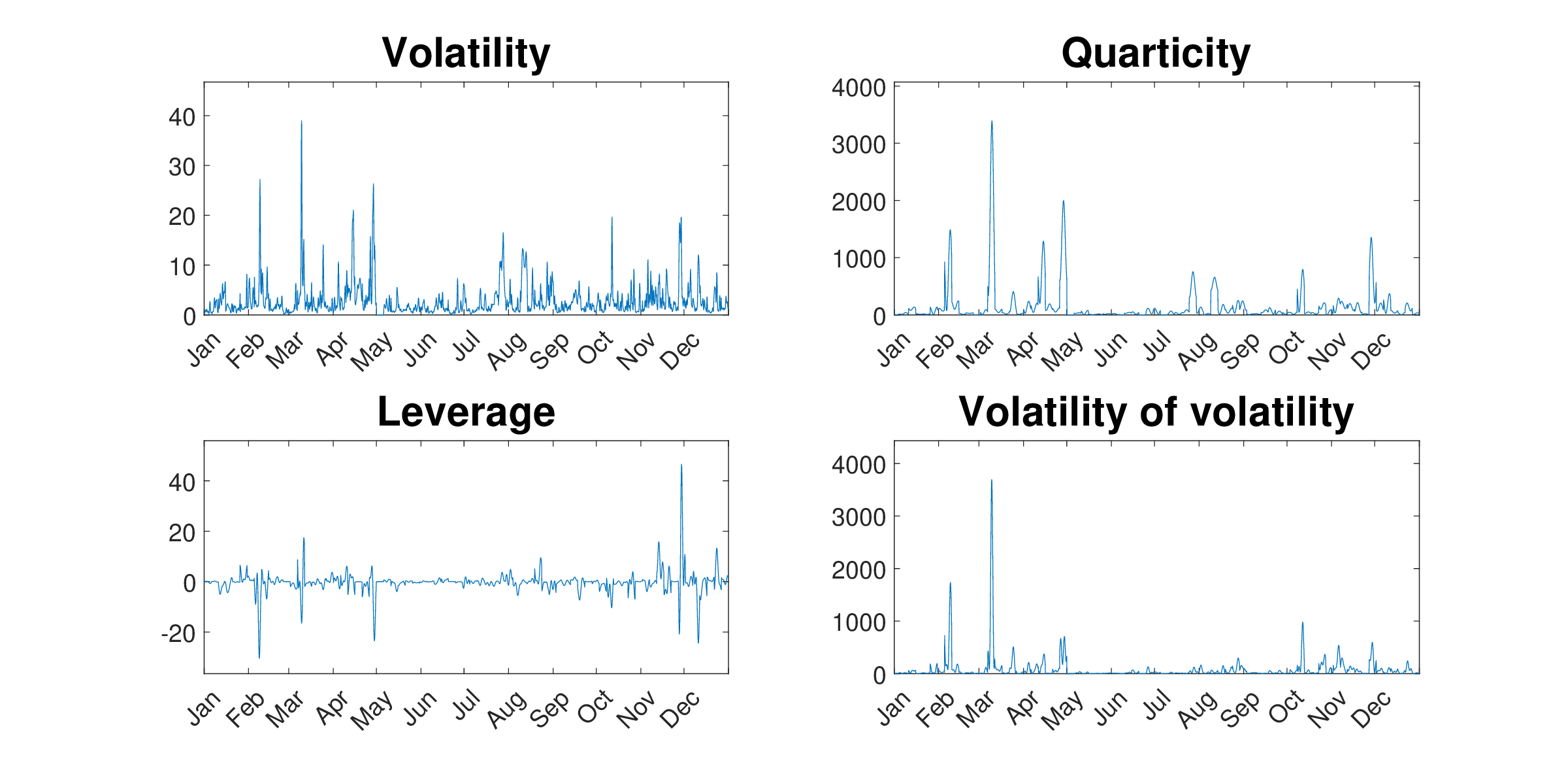}
        \caption{Estimated atmospheric temperature spot volatility, quarticity, leverage and volatility of volatility
for the geographic coordinates $-34.0, 151.1$  over 2022.}
        \label{fig:est_weather1}
    \end{figure}

Note that the choice of \verb|tau| is such that estimates are obtained every $4$ hours. The resulting sequence of spot estimates is displayed in Figure \ref{fig:est_weather1}, while Table \ref{tab:est_weather1} contains the aggregate values of integrated quantities, obtained as the sum of sub-sample estimates. Figure \ref{fig:est_weather1} shows that the spot volatility, quarticity and volatility of volatility show clusters, displaying larger spikes between February and May and smaller spikes near the end of the year. As for the spot leverage\footnote{We refer to the quadratic covariation between atmospheric temperature and its volatility as $leverage$ to underline that it has been estimated via the homonymous function of the FMVol library. However, we are aware that this is an abuse of terminology when input time series are not related to asset prices.}, Figure \ref{fig:est_weather1} shows a value close to zero on average, with significant negative and positive spikes occurring approximately in correspondence with spot volatility spikes. 

     \begin{table}[h!]
  \centering
  \begin{tabular}{ llll }
     {Volatility} &  {Quarticity} &  {Vol. of vol.} & {Leverage} \\ 
     \hline
     $   1060.552$ & $  11150.248 $ & $  18816.550 $   & $-89.279$  \\
  \end{tabular}
    \caption{Estimates of integrated quantities for the atmospheric temperature series corresponding to the geographic coordinates $-34.0, 151.1$  during 2022.}\label{tab:est_weather1}

\end{table}

\begin{figure}[h!]
        \centering
        \includegraphics[scale=0.2]{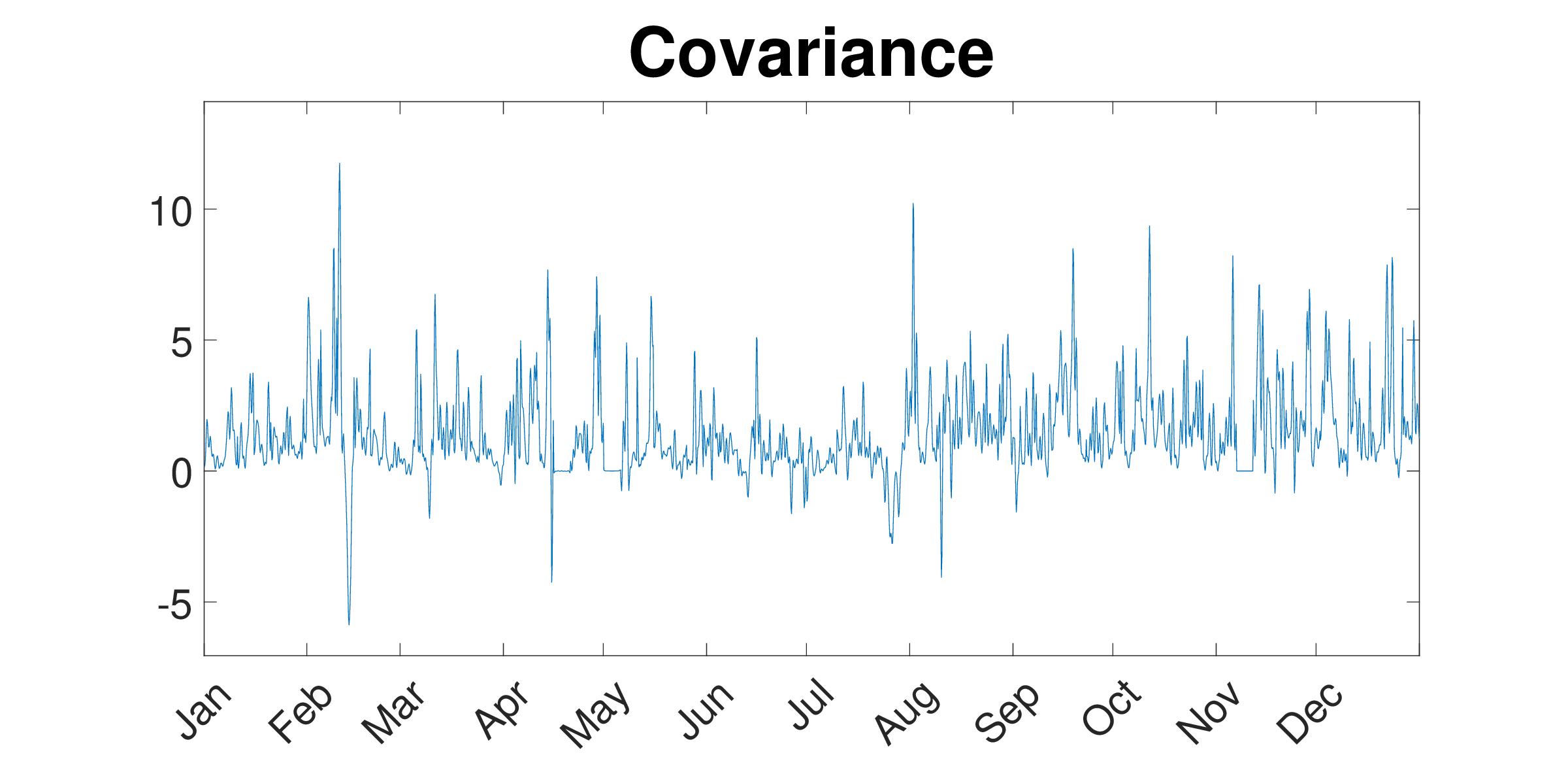}
        \caption{Estimated spot covariance of the atmospheric temperature recorded at coordinates $-34.0, 151.1$ and $-33.7, 150.8$ during 2022.}\label{fig:est_weather2}
    \end{figure}

We also estimate the spot and integrated covariance with de-trended residuals of the temperature time series recorded from the weather station with coordinates $-33.7, 150.8$, which has also been downloaded from the Copernicus database. Since the two series are equally spaced and synchronous, for the estimation of the integrated and spot covariance we maintain the same values of \verb|T|, \verb|N|, \verb|M|, \verb|L| and \verb|tau| as in the univariate case. The resulting spot estimates are displayed in Figure \ref{fig:est_weather2}. The aggregate integrated covariance amounts to $536.253$. We note that the spot covariance tends to be most often positive, consistently with the fact that the two locations are quite close in terms of geographic coordinates, with occasional negative spikes.

\section{Conclusions}\label{sec:concl}

This paper presents the Fourier-Malliavin Volatility (FMVol) estimation library for MATLAB$^{\circledR}$, which includes functions that implement the Fourier-Malliavin estimators of the spot and integrated volatility, co-volatility, quarticity, volatility of volatility and leverage. 

The paper aims to facilitate the implementation of Fourier-Malliavin estimators even for users who are not familiar with Fourier analysis, by providing a freely available library, in view of the flexibility of the method and the related wide scope of potential applications.

Although originally introduced for the estimation of asset volatilities, the Fourier-Malliavin method is indeed a general non-parametric method that can be applied whenever one is interested in reconstructing the latent volatility (and/or second-order quantities) of a stochastic volatility process from discrete observations thereof. We remark that the method, which relies on in-fill asymptotics, does not impose any stationarity and ergodicity assumptions. Further, we stress that the method does not require equally-spaced observations and is robust to measurement errors, or noise, without any preliminary bias correction or data pre-treatment. Moreover, in its multivariate version, it is intrinsically robust to irregular and asynchronous sampling.

\section*{Funding}

S. Sanfelici is a member of the INdAM-GNCS research group, Italy. 
G. Toscano gratefully acknowledges financial support from the Institut Louis Bachelier 2022
grant ‘Risk management in times of unprecedented geo-political volatility: a machine learning
approach’.

\section*{Acknowledgments}
The authors are thankful to Marco Riani and Aldo Corbellini for their support in developing the FMVol library.

\section*{Disclosure statement}
The authors have no competing interests to disclose.

\section*{Appendix. Bivariate data simulation} 

This Appendix describes the functions \verb|Heston2D.m| and \verb|Heston1D.m| that we used to simulate data samples in Section \ref{sect:toolbox}. See  \cite{Heston93} for more details on the parametric model employed.


The function \verb|Heston2D.m| implements the Euler scheme to integrate the stochastic differential equation (\ref{Heston1})-(\ref{Heston4}). The function reads as 
\begin{verbatim}
[x,V] = Heston2D(T,n,parameters,Rho,x0,V0), 
\end{verbatim}
where:
\begin{itemize}
    \item[-] \verb|T| (e.g., \verb|T=1|) denotes the length of the integration interval $[0,T]$;
    \item[-] \verb|n| (e.g., \verb|n=10000|) is the number of sub-intervals used for the discretization of the equations;
    \item[-] \verb|parameters| is $4 \times 2$ matrix whose $j$-th column contains the values for the parameters $\mu_j$, $\alpha_j$, $\theta_j$ and $\gamma_j$, $j=1,2$,  in (\ref{Heston1})-(\ref{Heston4});
    \item[-] \verb|Rho| is a column vector of length equal to $6$ that contains the correlation coefficients of the Brownian motions, namely $\rho_{1,2}$, $\rho_{1,3}$, $\rho_{1,4}$, $\rho_{2,3}$, $\rho_{2,4}$, $\rho_{3,4}$;
    \item[-] \verb|x0| and \verb|V0| are vectors that contain the initial values for $x^1, x^2$ and their related variances, respectively.
\end{itemize}
The function provides as outputs the $(\verb|n|+1) \times 2$ matrices \verb|x| and \verb|V| containing the $n+1$ observations for $x^1$ and $x^2$ and their related instantaneous variance observations, respectively.

The function \begin{verbatim}[x,V]=Heston1D(T,n,parameters,rho,x0,V0)\end{verbatim}
similarly generates a univariate diffusion process and its instantaneous variance from the Heston model.

We refer the reader to the related MATLAB$^{\circledR}$ documentation and help files available in the FSDA Toolbox for more details.
 
 \newpage
 \bibliographystyle{apalike}
\bibliography{FMbiblio}

\end{document}